\documentclass[fleqn,usenatbib]{mnras}
\usepackage{newtxtext,newtxmath}
\usepackage[T1]{fontenc}
\usepackage{graphicx}	
\usepackage{amsmath}	
\usepackage{hyperref}

\title[What determines the structure of sGRB jets?]{What determines the structure of short gamma-ray burst jets?}

\author[Urrutia et al.]{
Gerardo Urrutia,$^{1}$\thanks{E-mail: gerardo.urrutia@nucleares.unam.mx}
Fabio De Colle,$^{1}$
Ariadna Murguia-Berthier$^{2,3}$
and Enrico Ramirez-Ruiz$^{2,3}$
\\
$^{1}$Instituto de Ciencias Nucleares, Universidad Nacional Aut{\'o}noma de M{\'e}xico, A. P. 70-543 04510 D.F. Mexico\\
$^{2}$Department of Astronomy and Astrophysics, University of California, Santa Cruz, CA 95064, USA\\
$^{3}$DARK, Niels Bohr Institute, University of Copenhagen, Blegdamsvej 17, 2100 Copenhagen, Denmark
}
\date{Accepted 2021 March 03. Received 2021 March 03; in original form 2020 November 13}
\pubyear{2021}

\begin{document}
\label{firstpage}
\pagerange{\pageref{firstpage}--\pageref{lastpage}}
\maketitle


\begin{abstract}
The discovery of GRB 170817A, the first unambiguous off-axis short gamma-ray burst arising from a neutron star merger, has challenged our understanding of the angular structure of relativistic jets.
Studies of the jet propagation usually assume that the jet is ejected from the central engine with a top-hat structure and its final structure, which determines the observed light curve and spectra, is primarily regulated by the interaction with the nearby environment. However, jets are expected to be produced with a structure that is more complex than a simple top-hat, as shown by global accretion simulations.
We present numerical simulations of short GRBs launched with a wide range of initial structures, durations and luminosities. We follow the jet interaction with the merger remnant wind and compute its final structure at distances $\gtrsim 10^{11}$~cm from the central engine.
We show that the final jet structure, as well as the resulting afterglow emission, depend strongly on the initial structure of the jet, its luminosity and duration. While the initial structure at the jet is preserved for long-lasting SGRBs, it is strongly modified for jets barely making their way through the wind.
This illustrates the importance of combining the results of global simulations with propagation studies in order to better predict the expected afterglow signatures from neutron star mergers. 
Structured jets provide a reasonable description of the GRB 170817A afterglow emission with an off-axis angle $\theta_{\rm obs} \approx 22.5^\circ$.
\end{abstract}

\begin{keywords}
relativistic processes --
radiation mechanisms: non-thermal --
methods: numerical --
gamma-ray burst: general --
stars: jets
\end{keywords}

\maketitle


\section{Introduction}\label{sec:intro}
Short gamma-ray bursts (sGRBs) are intense flashes of $\gamma$-rays lasting $\lesssim 2$~s. They are produced by the coalescence of neutron stars, accompanied by the ejection of powerful relativistic jets with energies $\approx 10^{51}\,$~erg and opening angles $\theta_j \lesssim 10^\circ$ \citep[see][for a review]{kumar15}.

The merger of a neutron star binary can produce hyper-massive neutron stars (HMNS) for  equations of state  that allow a maximum mass for non-rotating neutron stars  in the range 2.6-2.8$M_\odot$ \citep{2006PhRvD..73f4027S,2008PhRvD..78h4033B}. Otherwise, the remnant will collapse to a black hole. In the former case, angular momentum transport from the inner to the outer regions of the remnant will drive significant mass loss through a neutrino-driven \citep{Rosswogetal2003,LeeRamirezRuiz2007,Dessart2009,Perego2014}, or a magnetically-driven wind \citep{Rezzolla2011,Siegel2014,Ciolfi2017,Ciolfi2020}. Different mechanisms responsible for the mass-loss naturally lead to different wind structures \citep[e.g,][]{Murguia-Berthier2017}.

As the jet traverses the wind environment, it decelerates depositing  a significant fraction of its  kinetic energy into an extended, hot cocoon \citep[e.g.,][]{ramirezruiz2002}. In this phase, the cocoon pressure helps to collimate the jet. Once the jet and cocoon break out of the dense environment (at a distance $\approx 10 ^{10}$ cm from the central engine), the cocoon expands laterally. The jet core, with an opening angle $\lesssim 5^\circ$--$10^\circ$, moves at highly relativistic speeds (with Lorentz factors $\Gamma\approx 100$), while the cocoon shows a steep velocity gradient in the polar direction, expanding initially at mildly relativistic speed ($\Gamma\approx$ 2-10) close to the core of the jet and at non-relativistic speeds near the equatorial plane \citep{kumar03,granot2005,Bromberg2011,Granot2012}.

The radial and angular structure of the jet and cocoon after the break-out regulates the late dynamics, and it is the key parameter in determining the shape of the light curve at large scales \citep[e.g.,][]{granot2005,duffell18,lazzati18,MooleyNature2018,Nakar2018} and the rate of off-axis sGRBs which will be observed by upcoming multi-wavelength surveys \citep{Salafiaetal2016,Ghirlanda2019,Gottlieb2019}. 

 As the GRB prompt and afterglow emission is strongly beamed and the relativistic jets are unresolved, the jet structure is poorly known. Observations of the GRB 170817A have given some insights on the angular structure of short GRBs. The afterglow emission has been interpreted as evidence of a structured jet (nevertheless, see \citealt{Gill2019} for an alternative explanation invoking a top-hat structure for the jet), moving  $\approx$ 20$^\circ$-30$^\circ$ off-axis with respect to the line of sight, with energy $E_{\rm iso} = 10^{49}-10^{51}$ ergs, ambient density $n_{\rm ISM}=10^{-4}-10^{-3}\,$cm$^{-3}$, and microphysical parameters $\epsilon_{e}\sim 10^{-2}$ and $\epsilon_{B}\sim 10^{-4}-10^{-3}$ \citep[where $\epsilon_B$ and $\epsilon_e$ are the fractions of post-shock thermal energy ending into the energy of the magnetic field and accelerated electrons, respectively; see, e.g.,][]{Murguia-Berthier2017,granot2018,lazzati18,MooleyNature2018,Nathanail2020}, although there is a large degeneracy in the determination of these parameters.

Numerical simulations of the jet propagating through the environment of the central engine have clarified the role that the wind plays in shaping the structure of the jet \citep{Aloy2005,Murguia-Berthier2014,Kathirgamaraju2018, Murguia-Berthier2017,Bromberg2018,duffell18,granot2018,lazzati18,Lamb2018,Xie2018,Gill2019,Lamb2019,Lazzati_2019,Lazzati_2020,Hamidani2020,Murguia-Berthier2020}.\footnote{The wind can also be dense enough to completely chocked the jet.}

Extrapolating these simulations up to distances $\approx 10^{16}\,$cm, one can solve the inverse problem and attempt to  constrain the small scale structure of the jet \citep[e.g.,][]{duffell18,lazzati18,MooleyNature2018,Nathanail2020}. 
Thus, the interaction of the jet with the environment and the resulting jet structure can in principle constrain the mechanism leading to both the mass-loss during the merger as well as the initial structure of the jet \citep{2002MNRAS.336L...7R, 2003MNRAS.343L..36R, Murguia-Berthier2014,Murguia-Berthier2017,Murguia-Berthier2020}.

Numerical simulations of sGRB typically assume that the jet is launched with a top-hat structure. On the other hand, numerical simulations of the jet formation \citep{RosswogLiebendorfer2003,Rosswogetal2003,Rezzolla2011,McKinney2012} show that the jet density and velocity structures are more complex than a simple top-hat.
For instance, GRMHD numerical simulations from \cite{Kathirgamaraju2019} show that jet is strongly structured at a distance of $\approx 6\times 10^8$~cm. Observations of AGN jets also indicate that jets are not homogeneous along the angular direction, with a low-density, fast moving material in the inner region of the jet associated to more dense and slower moving material at large polar angles \citep[see, e.g.,][and references therein]{Walg2013}.

The role played by the intrinsic jet properties (jet opening angle, structure, luminosity, duration and  magnetic field at the launching point) in determining the final jet structure has been not been studied yet in detail. Thus, in this paper we present numerical simulations of jets structured along the polar direction, i.e., with a jet luminosity $L_j(\theta)$ and Lorentz factor $\Gamma(\theta)$, and compare the outcome with simulations that assume a top-hat jet. 
We also explore the effect of changing the injection duration of the jet $t_j$ and the luminosity history $L(t)$, which is expected to decrease with time, as the disk around the merger remnant is viciously drained \citep[e.g.,][]{lee05}.

This paper is structured as follows: in section \ref{sec:methods} we describe the code and the initial conditions employed in our simulations. In section \ref{sec:results} we present the results of the numerical simulations and the final structure of top-hat and (intrinsic) structured jets. In section \ref{sec:discussion} we discuss the results and their observational implications. Finally, in section \ref{sec:conclusions} we summarize our results.

\section{Methods}\label{sec:methods}

\begin{table}
\centering
\caption{Initial conditions of the numerical simulations.}
\begin{tabular}{|c||c|c|c||}
\hline
Model & $t_{\rm j}$ & $L_{\rm j}$      & jet structure    \\
& (s) & (erg s$^{-1}$) &     \\ \hline \hline
&&& Top-hat \\
Low Luminosity TB  & 0.6    & $1.5\times 10^{49}$ & Gaussian \\
&&& Power-law \\ \hline
&&& Top-hat \\
Low Luminosity TM  & 1.5    & $1.5\times 10^{49}$ & Gaussian \\
                   &             &                     & Power-law      \\ \hline
&&& Top-hat \\
Low Luminosity TL  & 3.0      & $1.5\times 10^{49}$ & Gaussian \\
                   &             &                     & Power-law      \\ \hline \hline
&&& Top-hat \\
High Luminosity TB & 0.5    & $1\times 10^{50}$   & Gaussian \\
                   &             &                     & Power-law      \\ \hline
                   &             &                     & Top-hat  \\
High Luminosity TM & 1.5    & $1\times 10^{50}$   & Gaussian \\
                   &             &                     & Power-law      \\ \hline
                   &             &                     & Top-hat  \\
High Luminosity TL &  3.0      & $1\times 10^{50}$   & Gaussian \\
                   &             &                     & Power-law      \\
 \hline \hline

Time-dependent Luminosity & 2.0 & $L_{\rm jet}(t)$ & Top-hat\\ \hline 
\end{tabular}
\label{tab1}
\end{table}

We study the propagation of sGRB jets through the pre-collapse merger remnant environment  by running a set of special relativistic hydrodynamics (SRHD) simulations. We use the adaptive mesh refinement (AMR) code \emph{Mezcal} \citep{decolle12}.
The code employs a second-order Runge-Kutta time integrator, and second-order space interpolation, which reduces to first order by a minmod limiter. We use the HLL method \citep[][]{Harten1983}. Coupling the HLL method with the minmod limiter makes the method very robust, although somehow dissipative.
In the relativistic version of the code, the primitive variables (the density $\rho$, the velocities $\vec{v}$ and the pressure $p$) are determined from the conservative variables ($D=\rho \gamma$, $\vec{m}=Dh\gamma \vec{v}, \tau=Dh\gamma c^2-p-Dc^2$, where $\gamma$ the Lorentz factor and $h$ the enthalpy) by using an iterative Newton-Raphson method for the equation $F(\Theta)= h(\Theta) \gamma(\Theta)-\Theta/\Gamma(\Theta)-1-\tau/Dc^2$, being $\Theta=P/\rho c^2$ (see \citealt{decolle12} for more details).

As detailed below, we explore the effect of changing the jet duration $t_j$, luminosity $L_j(\theta, t)$ (as a function of time and polar angle $\theta$) and jet angular structure, i.e., the jet density $\rho_j(\theta)$ and velocity $\Gamma_j(\theta)$.\footnote{The role played by the wind duration $t_w$ and structure has been explored in detail by \cite{Murguia-Berthier2014, Murguia-Berthier2017,Murguia-Berthier2020} and is not considered here. We also do not consider the jet magnetic field, which can also play an important role in determining the jet structure \citep{duffell18,gottlieb2020b,Nathanail2020}. } 
The initial conditions of the simulations presented in this paper are listed in Table \ref{tab1}.

The simulations are done by using a two-dimensional axisymmetric grid, with $800 \times 800$ cells along the $r$ and $z$ direction at the lowest level of refinement and with 6 levels of refinement,  corresponding to $25600\times 25600$ cells at the highest level of refinement. The size of the computational box is $(L_r,L_z)=(4.8, 4.8) \times 10^{11}\,$cm, with a maximum resolution of $1.8 \times 10^{7}\,$cm.

At $t=0$, the computation box is filled with a static medium with density $\rho_a=n_a m_p$ (with $n_a=10^{-5}$ cm$^{-3}$) and a pressure $p_a =10^{-10} \rho_a c^2$. As the jet and wind are highly supersonic, the results are independent of the pressure of the ambient medium.
From an inner boundary located at $r_{\rm in}=10^9$ cm, we inject a wind, with a velocity $v_w=c/3$ and a wind mass-loss $\dot{M}_w = 10^{-4}$ M$_\odot$ s$^{-1}$ during a time $t_w = 1$ s.
At $t>t_w$, the wind density is gradually switched-off by dropping the density as $\rho_w\propto t^{-5/3}$.
These wind mass-loss and duration are typical of NS mergers \citep{Qian_1996,rosswogRamirezRuiz2002,LeeRamirezRuiz2007}. The effect of more dense ($\dot{M} \gtrsim 10^{-3}\,M_{\odot}$) and a spherical wind structures \citep[e.g.,][]{Perego2014} were considered by \cite{Murguia-Berthier2014,Murguia-Berthier2020}.

\begin{figure}
    \centering
    \includegraphics[scale=0.35]{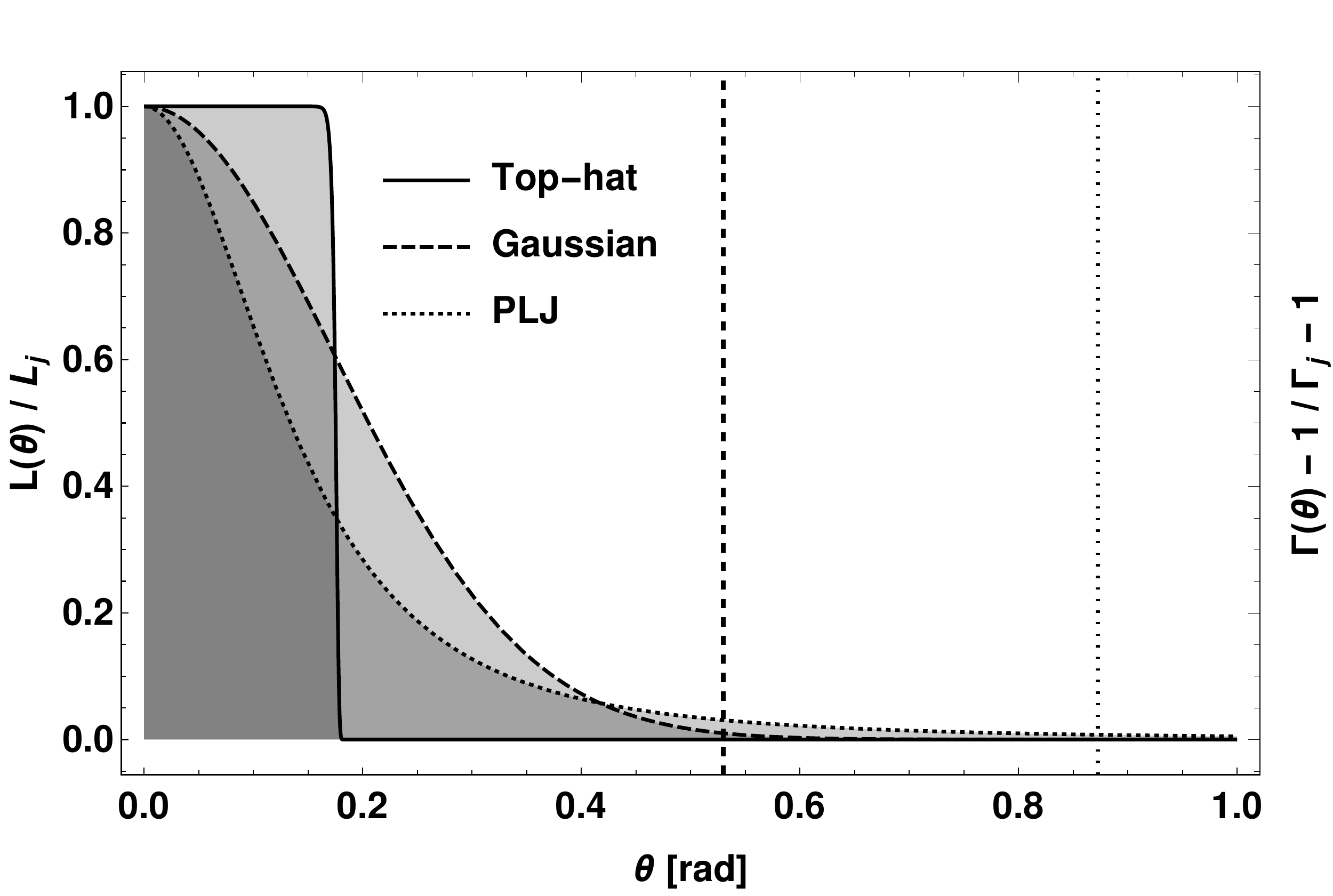}
    \caption{Kinetic luminosity and Lorentz factor profiles for the structured jets implemented in the numerical simulations. The top-hat jet (solid line) has a cut-off angle of $\theta_j=10^\circ$. Gaussian (dashed line) and power-law jet (dotted line) profiles present extended wings moving with Newtonian velocity.}
    \label{fig1}
\end{figure}

Starting at $t=t_w$, and during a time $t_j$, a jet is launched from a radius $r_j(\theta) = r_w v_j(\theta)/v_{j0}$ (hereafter, the sub-index $0$ indicates values computed at $\theta=0)$. 
We implement three different structures for the jet (see Figure \ref{fig1}), changing its luminosity and velocity as a function of the polar angle. The jet structure is defined as
\begin{eqnarray}
  f(\theta)=\left\{ \begin{array}{cc} 
  1 & \qquad \theta \leq \theta_j \\
  0 & \qquad \theta>\theta_j
  \end{array} \right. \qquad &&{\rm ``Top\; hat"\; jet\;,} \\
    f(\theta) = e^{-\frac{1}{2} \frac{\theta^2 }{\theta_j^2 } } \qquad &&{\rm ``Gaussian"\; jet\;,} \\
 f(\theta) = \left( 1+\frac{\theta^2}{\theta_j^2}\right)^{-3/2}
 \qquad &&{\rm ``Power \; law"\; jet\;,}
\end{eqnarray}
being $\theta_j=0.2$ rad the jet core, in which most of the energy is injected.

The jet has a kinetic luminosity 
\begin{equation}
  L_j(\theta) =  L_{j0} f(\theta),
\end{equation}
and a Lorentz factor 
\begin{equation}
  \Gamma_j(\theta)= 1 + (\Gamma_{j0}-1) f(\theta) \;,
\end{equation}
where $\Gamma_{j0}=20$. The jet density is defined as a function of luminosity and Lorentz factor as
\begin{equation}
    \rho_j(\theta) = \frac{L_{\rm iso}(\theta)}{4\pi r_j(\theta)^2 \Gamma_j^2(\theta) c^3}\;,
\end{equation}
where the isotropic luminosity is related to the jet luminosity by $L_{j} = L_{\rm iso} \left(1-\cos \theta_j \right)\approx L_{\rm iso}\theta_j^2/2$.
We assume that the jet is cold, by setting its pressure as $P_j=10^{-5} \rho_j c^2$.
We run simulations by employing typical sGRBs luminosities \citep[e.g.,][]{Levan2016}, i.e.  $L_j = 10^{50}$ erg s$^{-1}$, but we also run simulations with lower jet luminosities
$L_{j} = 1.5 \times 10^{49}\,$ erg s$^{-1}$. 

We run simulations for different jet lifetimes $t_j$ (see Table \ref{tab1}). The time $t_j=$ TB corresponds approximately to the break-out time ($t_{\rm bo}$), i.e. $t_j=0.5$ s and $t_j=0.6$ s for the high and low luminosity top-hat model (similar values are obtained also for structured jets as the core energy is similar in all our models).
The time $t_j=$ TM is taken by considering the typical duration of a sGRB $T_{\rm 90} \sim 0.9\,$s \citep[e.g.,][]{Berger2014, Levan2016}  and $t_j = t_{\rm bo} + t_{\rm B}$. We also run simulations by using $t_j$ = TL $ = 3\,$ s, corresponding to sGRB in the tail of the duration distribution.

In addition, we run simulations with a top-hat jet structure considering a luminosity that changes with time.
We consider a constant luminosity followed by a power-law decrease,
\begin{equation}
  L(t)=\left\{ \begin{array}{cc} 
  L_0 & t<t_0 \\
  L_0 (t/t_0)^{-5/3} & t>t_0 \end{array} \right.
  \label{eqn:hist_AC}
\end{equation}
which reduces to $L_0=E_{\rm j}/t_{\rm j}$ when $t_0=t_j$. We take $t_0 = t_j/10$. 

Finally, we caution that two-dimensional simulations are prone to a numerical instability known as the  ``plug'' instability \citep{lazzati15}. This instability results from the symmetry used in two-dimensional simulations, and disappears in three-dimensional simulations once the symmetry with respect to the jet axis is broken by taking asymmetric initial conditions (e.g., a small jet precession or asymmetric variations in the ambient medium) or by employing asymmetric numerical methods in the integration of the SRHD equations\footnote{When using unsplit solvers and symmetric initial conditions, three dimensional numerical simulations are identical (at machine precision) to two dimensional simulations.} (e.g., split solvers). In Appendix A, we show how this instabilities can be suppressed by slightly changing the jet velocity direction.

In three-dimensional simulations, instabilities generated at the contact discontinuity extend to the jet axis, mixing baryon-rich material with the baryon poor jet material. The baryonic contamination then reduces the jet velocity \citep[e.g.][]{gottlieb2020,Gottlieb2020a,harrison18}.
However, this happens only in dense media, so it affects more LGRBs than SGRBs (which cross a few \% of the material than LGRBs). 
Also (as we will show in the next section), in structured jets the contact discontinuity is pushed away from the jet axis, then dropping the amount of baryon-rich material reaching the jet.

\section{Results}\label{sec:results}
\subsection{Jet dynamics}

\begin{figure*}
    \centering
    \includegraphics[scale=0.185]{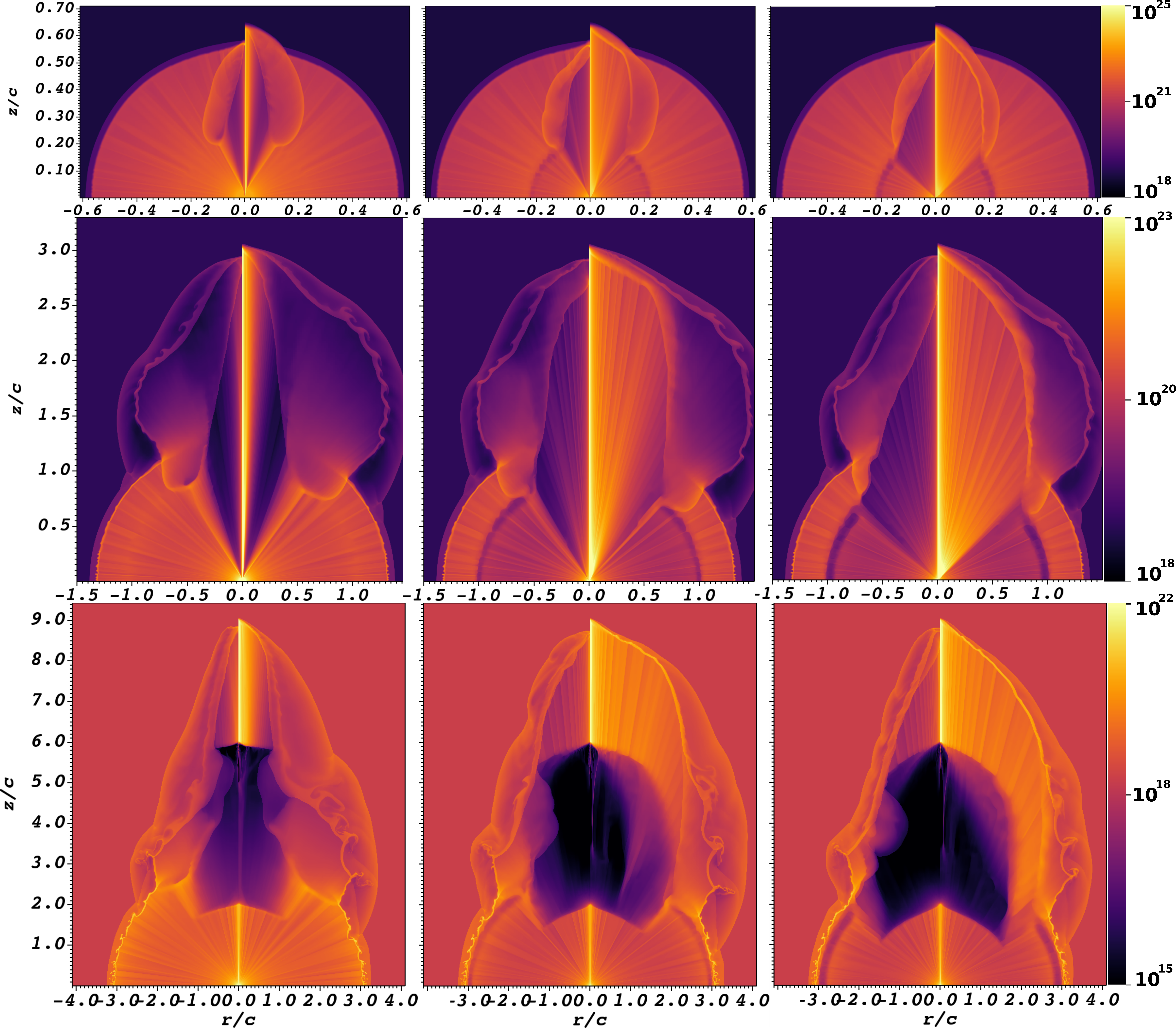}
    \caption{Number density maps for jets which time of injection is $t_j=3\,$s. \emph{Left to right panels:} top-hat, gaussian and power-law jet. In each panel, the left (right) half shows the low (high) luminosity jet.  \emph{Top to bottom panels:} snapshots at the breakout time $t=1.6$ s (or $t-t_w=0.6$ s); 4 s (or $t-t_w=3\,$s), when the jet is switched off; and 10 s (or $t-t_w=9\,$s), corresponding to the final time of the simulations.}
    \label{fig2}
\end{figure*}


\begin{figure*}
    \centering
    \includegraphics[scale=0.22]{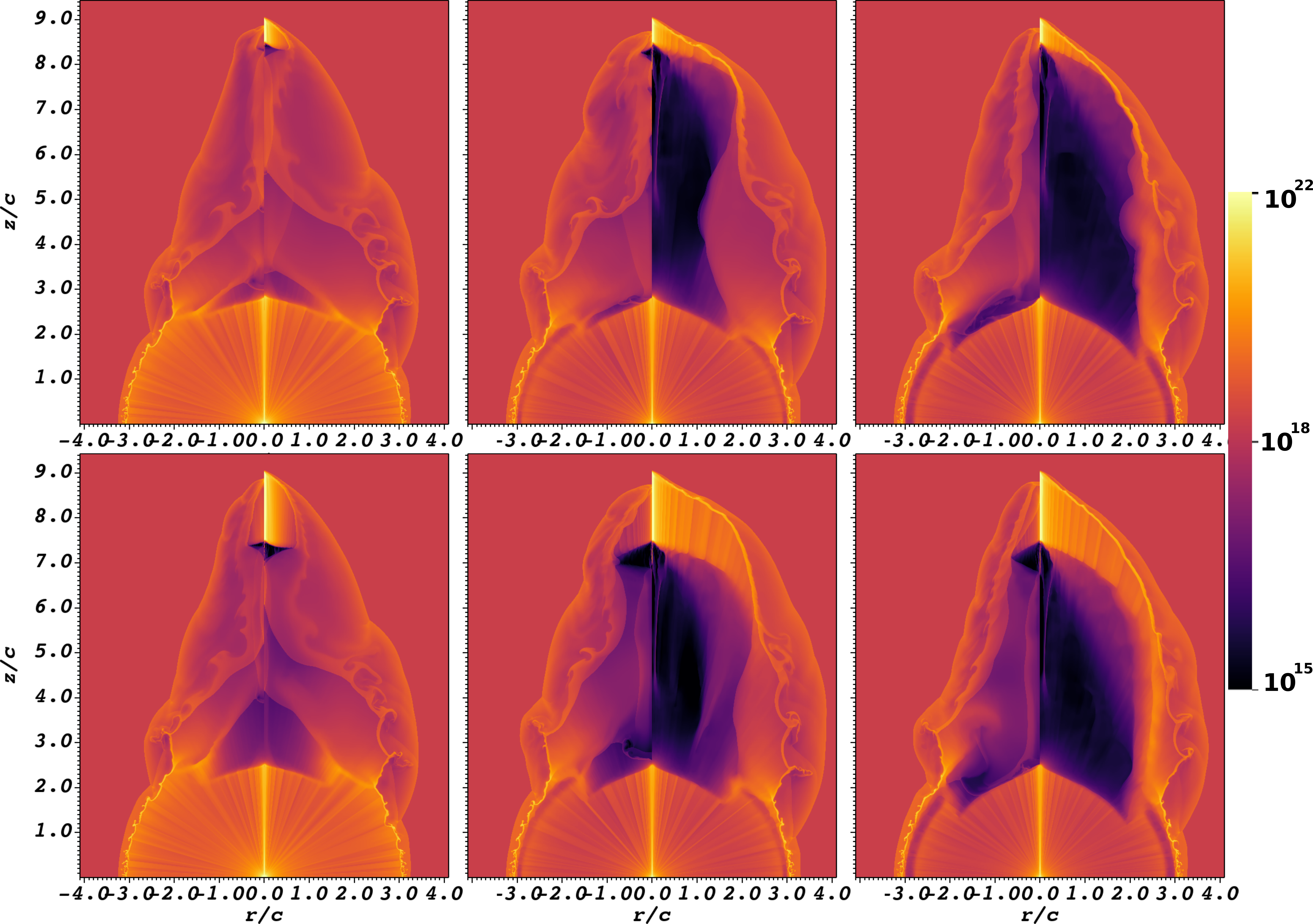}
\caption{Number density maps for jets injected during a time $t_j$, seen at $t=10$ s. \emph{Left to right panels:} top-hat, gaussian and power-law jet. \emph{Upper panels}: models with $t_j=0.6/0.5\,$s (each panel shows low $L_j$ in the left half and high $L_j$ in the right half). \emph{Bottom panels}: models with jets lasting $t_j=1.5\,$s. Similar figures for the models with $t_j=3\,$s are shown in the bottom panels of figure \ref{fig2}.}
    \label{fig3}
\end{figure*}

Figure \ref{fig2} shows the evolution of top-hat, Gaussian and power-law jets (from left to right in the figure) at time $t=1.6$ s, $t=4$ s and $t=10$ s (top to bottom panels). Each panel is divided into two halves, with the left (right) half showing the low (high) luminosity case. All panels refer to a jet injected during $t_j=3$ s (TL case, see Table \ref{tab1}).

The propagation of a top-hat jet through the environment has been considered by several authors \citep[e.g.,][]{Morsony2007,lopezcamara2013,Murguia-Berthier2014,lazzati15,Duffell2015,lopezcamara2016,Murguia-Berthier2017,decolle18a,decolle18b,lazzati18,duffell18,harrison18,Nathanail2020,Hamidani2020,Murguia-Berthier2020}. 
The shock velocity can be estimated by assuming ram pressure balance \citep[e.g.,][]{Begelman1989, ramirezruiz2002, Bromberg2011, decolle12b}, and is given as $v_{\rm sh} = v_j/(1+(\rho_a/\rho_j \Gamma_j^2)^{1/2})$, where $\rho_a\propto r^{-2}$ and $\rho_j(\theta) \propto L_j(\theta)$ the ambient and jet density respectively.
The wind and the jet are injected from a radial distance $r_w = 10^9$ cm from the central engine. At this radius, on axis, $\rho_a \ll \rho_j \Gamma_j^2$, then the jet head moves with a velocity $v_{\rm sh} \approx v_j$ in the high luminosity, and $v_{\rm sh} \lesssim 0.9 \; v_j$ in the low luminosity case. As $L_j(\theta)$ quickly drops with angle (see Figure \ref{fig1}), the wide angle jet component present in structured jets (middle, right top panels) moves at lower speeds.

As the jet transverses the wind medium, the hot shocked gas forms an extended cocoon (figure \ref{fig2}, top panels) which expands at mildly relativistic speeds. The heated plasma is made by the dense, shocked wind, and by the rarefied, shocked jet (the dark yellow region close to the jet axis in figure \ref{fig2}) separated by a contact discontinuity. The cocoon pressure helps to collimate the jet, which acquires a nearly cylindrical shape \citep[see, e.g.,][]{ramirezruiz2002,Bromberg2011,Duffell2015}.

As the jet core luminosity (i.e., the region corresponding to $\theta\lesssim \theta_j$) is approximately the same for all jet structures, the velocity of the jet as well as the break-out time are nearly independent on the jet structure but depend strongly on its luminosity. 
The low luminosity jet reaches the wind shock front at $t=1.6$ s (or $t-t_w=0.6$),  while high luminosity jets reach the boundary at $t-t_w=0.5$ (see the upper panel of Figure \ref{fig2}). 

The jet structure at the launching point strongly affects the cocoon structure. 
The wide angle outflow component expands at lower speeds, pushing and compressing the cocoon towards larger radii. Differences between different structures are more evident for high luminosity than for low luminosity jets (see Figure \ref{fig2}). 

The middle and bottom panels of Figure \ref{fig2} show the evolution of the jet as it breaks out of the wind and propagates into the diluted environment. As soon as the jet breaks out of the dense wind, the cocoon quickly expands sideways (at the post-shock sound speed, i.e. $v\approx c/\sqrt{3}$) engulfing the wind medium.
Gaussian and power-law jet cocoons expand laterally faster than the top-hat cocoon, as they are pushed from behind by the wide-angle outflow. Structured jets have also different cocoon structures, specially at intermediate angles in which Gaussian jets are more energetic than power-law jets.

The final jet structure is acquired a few seconds after the jet had drilled through the medium launched during the neutron star merger \citep{LeeRamirezRuiz2007,Murguia-Berthier2014,Murguia-Berthier2017,Murguia-Berthier2020,Duffell2015,Bromberg2018}. 
The final snapshot of our simulations is shown in the bottom panel of figure \ref{fig2}. As the jet is switched off, a low-density cavity\footnote{ Low-density, near-vacuum regions are often numerical challenging. The simulations  are done by employing a second-order (quite dissipative) HLL integrator. When the integrator fails to find a physically acceptable solution, the code switches to first order.} is formed behind the relativistic shell (with a size of $\lesssim 3$ light seconds).

Figure \ref{fig3} shows density maps at $t=10$ s as a function of the jet duration, structure and luminosity. As shown in Figure \ref{fig2}, the cocoon expand faster in structured jets. The cocoon shape depends on the amount of energy deposited by the jet inside the dense wind. Once the jet breaks out of it, the energy injected from the central engine ends into the jet core only. Then, jets with longer injection times (bottom vs. top panels panels) forms a more elongated and energetic shell of fast moving material, while the cocoon structure remains  nearly independent on the jet duration.

The perturbations extending radially through the wind (see, e.g., the bottom left panel of Figure \ref{fig3}) are a numerical artifact resulting from injecting a ballistic (i.e., with a kinetic energy much larger than its thermal energy) spherical wind into non-spherical cylindrical coordinates. These small perturbations do not affect the jet propagation.

\subsection{Jet structure}

\begin{figure}
    \centering
    \includegraphics[scale=0.26]{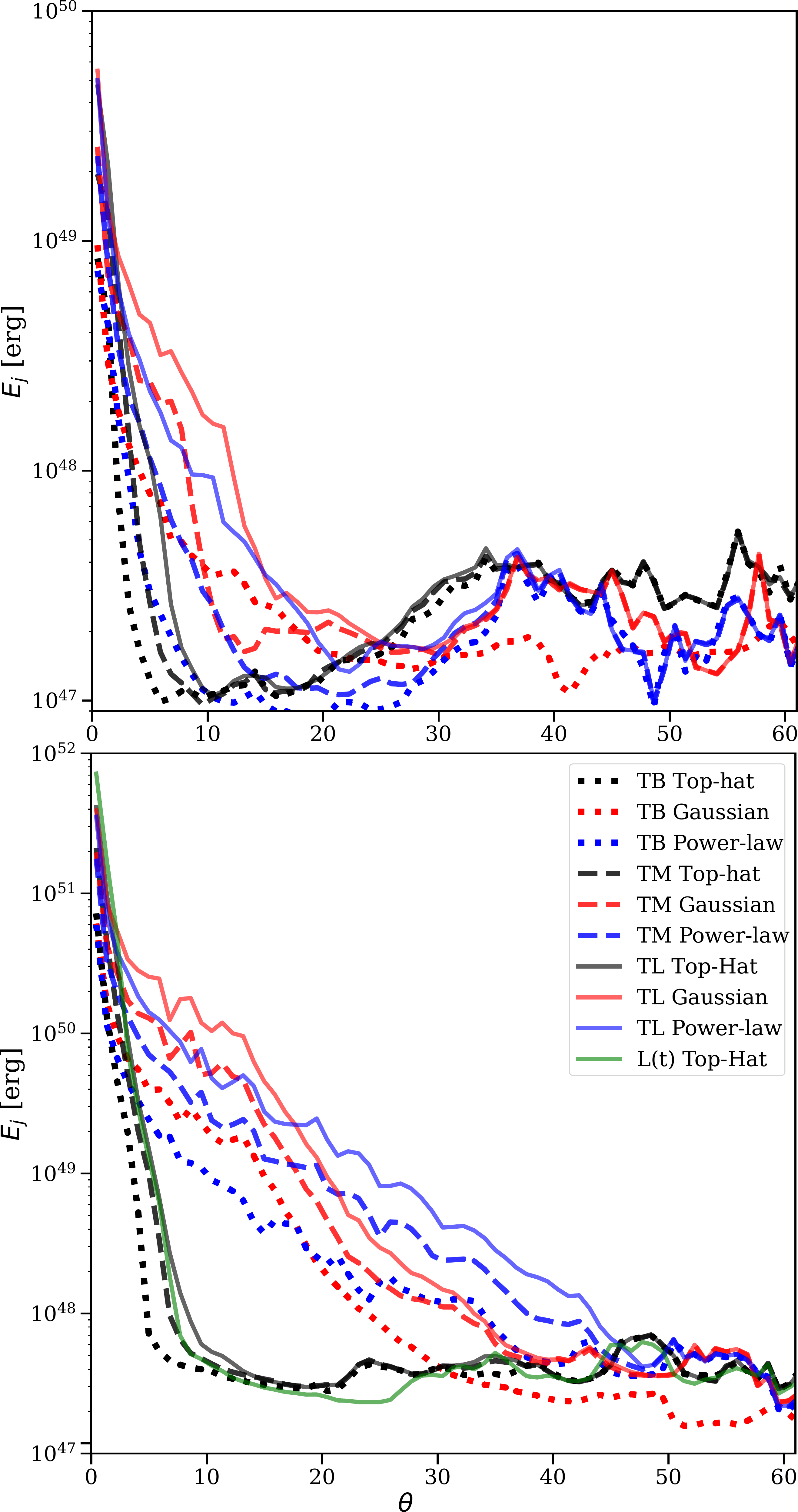}
\caption{Energy of the jet/cocoon system as a function of  $\theta$ (measured in the lab frame) at $t=10$ s. \emph{Top:} low luminosity models. \emph{Bottom:} high luminosity models. Models TB, TM, TL correspond to $t_j=0.5$ s, 1.5 s, 3 s respectively. Model $L(t)$ (bottom panel) corresponds to a jet with a luminosity changing with time.}
    \label{fig4}
\end{figure}

Once the jet/cocoon system breaks out of the wind medium, it expands and accelerates to highly (the jet) and mildly (the cocoon) relativistic speeds. Then, the jet/cocoon system  will expand freely up to distances $R\approx 10^{16}-10^{17}$ cm before decelerating. Thus, the structure of the jet coming out from our simulations will determine the structure of the jet and, as such, the afterglow emission, specially in SGRBs seen off-axis \citep[e.g.,][]{Murguia-Berthier2014,Murguia-Berthier2017,2017ApJ...848L..34M,lazzati18,Nakar2018,Irwin2019,Murguia-Berthier2020}.

Figure \ref{fig4} shows the energy at $t=10$ s, as a function of  polar angle $\theta$. Each panel illustrates the dependence of $E_j$ on the jet launching structure and duration. Here $E_j$ includes the cocoon's contribution. Comparing the upper and bottom panels one can also see the effect of increasing the jet's luminosity.
For all the models, most of the energy continues to be concentrated  into the jet's core.

In all models, the increase in the jet's duration leads to an increase of $E_j$  (which is proportional to the jet duration) and this can be as large as a  factor of $\approx$ a few at angles $\gtrsim 10^\circ$.
In structured, low-luminosity jets the energy deposited outside the jet core ($\approx  5^\circ-20^\circ$) is larger than in top-hat jets by $\approx$ an order of magnitude, with the Gaussian jets being more energetic than the power-law jets. Differences between top-hat and structured jets increase dramatically for high luminosity jets (see Figure \ref{fig4}, bottom panel). While in top-hat jets most of the increase in the luminosity resides in the jet's core, in structured jets leads to a large spread in energy up to angles $\lesssim 45^\circ$. Consistent with their initial energy distribution (see Figure \ref{fig1}), Gaussian jets are slightly more energetic than power-law jets at intermediate angles, and less energetic at large angles (although the exact initial jet structure is certainly not preserved).
The bottom panel of Figure \ref{fig4} shows also that a jet with a time-variable, top-hat luminosity has a final structure nearly identical to a top-hat jet with constant luminosity. Yet, it is more energetic on-axis given its higher total power.

These results can be understood by taking into account that, as the jet moves through the wind, it dissipates part of its kinetic energy, effectively transferring energy from the collimated jet to the uncollimated cocoon. Once the jet has broken out of the wind, the energy travelling into the jet channel remains mostly collimated within  the initial jet opening angle. The jet luminosity, opening angle, duration and the density structure of the environment all together determine the jet break out time and the amount of energy deposited into the cocoon.
Thus, the final jet structure will tend to mimic the initial shape at injection as the jet duration increases. We conclude that on average, the structure of long-lasting SGRBs tracks the initial structure at the jet (at the launching point), while the structure of SGRBs which just barely make  their way through the wind are significantly more affected by the external density stratification.

\section{Discussion}\label{sec:discussion}
The cocoon is important for  the following reasons. First, the cocoon pressure helps collimate the jet, which then acquires a nearly cylindrical shape. This has been extensively studied in the literature \citep[see, e.g.,][]{ramirezruiz2002,Bromberg2011,Duffell2015}. Second,  once the cocoon breaks out of the wind, its structure determines the large scale angular structure of the jet. In what follows, we discuss briefly the role played by the initial jet structure in 
regulating the cocoon energy structure\footnote{Jet rotation can also affect the cocoon structure, as shown in the context of AGN jets \citep[e.g.,][]{Keppens2008, Walg2013}. }. 

The structure of the pre-merger environment of the NS merger, shaped by the pre-collapse remnant  wind  \citep{LeeRamirezRuiz2007, Duffell2015} is an important component in determining the jet structure \citep{Murguia-Berthier2014, Murguia-Berthier2017, duffell18, Nakar2018, Irwin2019,Murguia-Berthier2020}. A non-homogeneous wind, in particular, can also drastically change the jet dynamics \citep[e.g.,][]{Murguia-Berthier2014, Murguia-Berthier2017,Murguia-Berthier2020}, facilitating the propagation of the jet along the polar regions.
Nevertheless, as shown in the previous section, the jet structure after breaking out of the wind depends mainly  on the initial jet structure. Depending on the break out time and the jet luminosity, cocoon energies for top-hat and structured jets can drastically differ.

\begin{figure}
    \centering
    \includegraphics[scale=0.35]{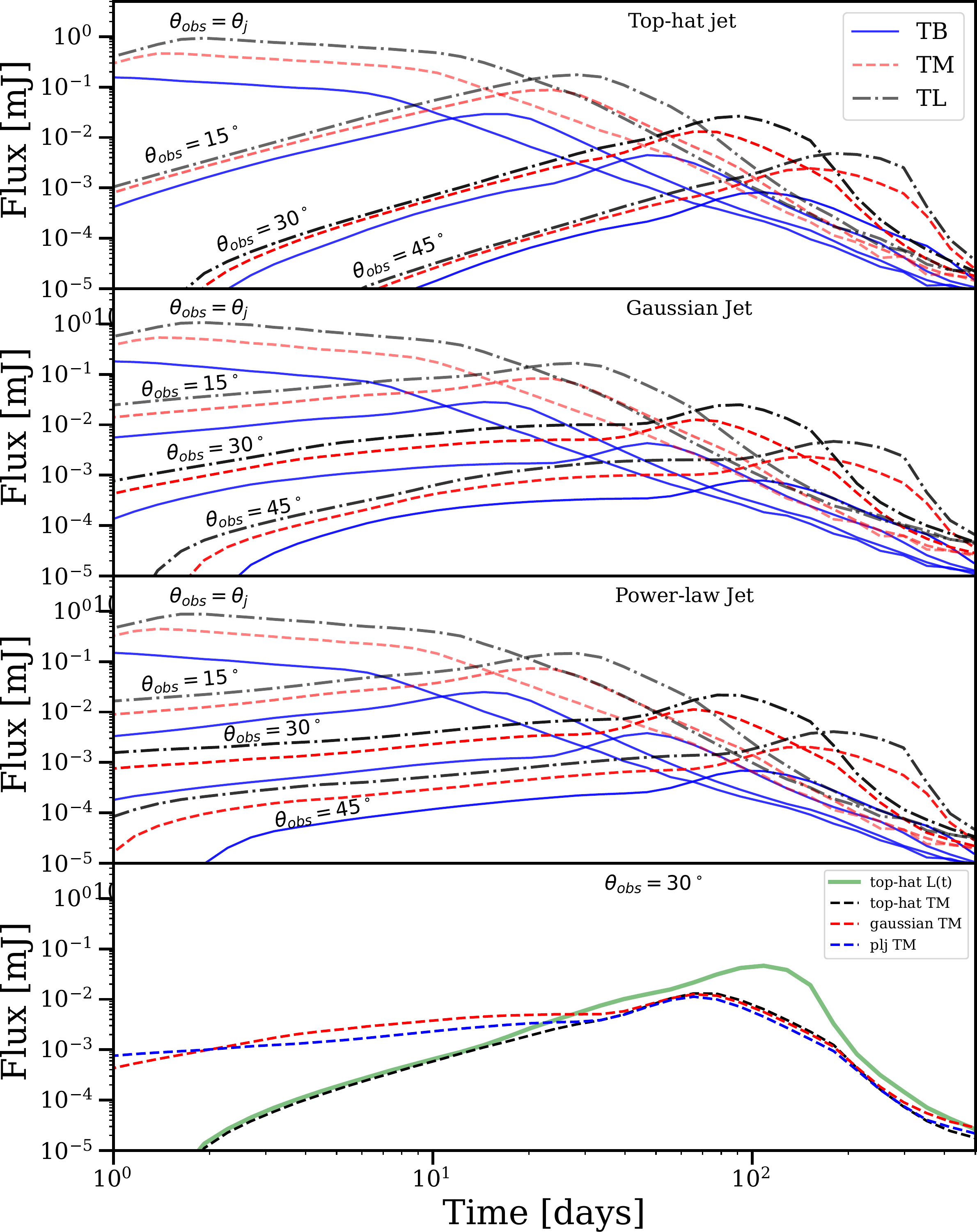}
    \caption{Light curves at $3\,$GHz for the top-hat, gaussian and power-law jet models (from top to bottom), as a function of observing angle. The bottom panel shows a direct comparison between the different jet structures at $\theta_{\rm obs} = 30^\circ$. The light curves have been computed by assuming a constant density stratification, with $n=1$ cm$^{-3}$, $\epsilon_e=0.1$, $\epsilon_B = 0.1$, $p=2.16$ and $D=40\,$Mpc. The jet luminosity is $10^{51}$ erg s$^{-1}$.}
    \label{fig5}
\end{figure}

\begin{figure}
    \centering
    \includegraphics[scale=0.32]{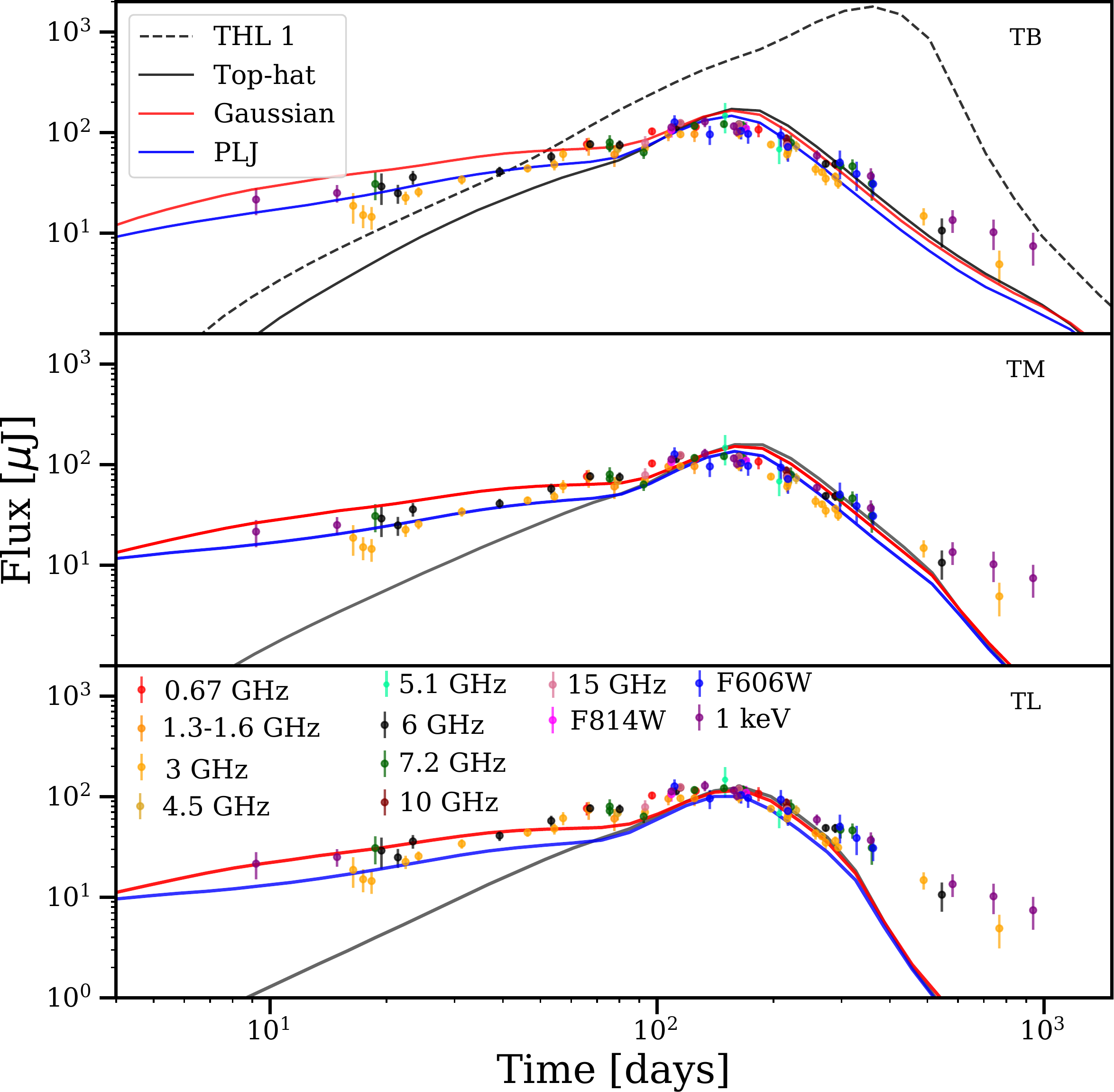}
    \caption{Comparison between observations of GRB 170817A and light curves computed by considering different jet structures viewed $\theta_{\rm obs}=22.5^\circ$. The light curves have been computed by assuming a constant density stratification, with $n=10^{-3}-10^{-2}$ cm$^{-3}$ (depending on the model), $\epsilon_e=0.01$, $\epsilon_B = 0.001$, $p=2.16$, and $D=40\,$Mpc. Observations are rescaled at 3 GHz \citep{Makhathini2020}. We stress that there is a large degeneration into the choice of the microphysical parameters (see the text for discussion).}
    \label{fig6}
\end{figure}

As mentioned before, the final jet structure after breakout has important consequences for the afterglow radiation, as clearly illustrated in the case of GRB 170817A  \citep{lazzati18,MooleyNature2018,Nathanail2020}. To determine how the jet's structure affects the SGRB emission, we computed the afterglow emission by extrapolating analytically the results of our numerical simulations up to distances\footnote{We note that we have assumed a constant density ISM, yet the structure of the medium at large radii might be altered if the binary NS hosts a pulsar \citep{holcomb14,2019ApJ...883L...6R}.} $\gtrsim 10^{18}$ cm. This is done by assuming synchrotron emission and considering a one-dimensional evolution at each polar angle, i.e. neglecting lateral expansion. 
The flux is computed by considering the contribution of each angular slice, which evolves independently in time (i.e., we neglect lateral expansion).
Given an energy $E_j(\theta)$ and an initial Lorentz factor $\Gamma_0(\theta)$, we assume that the shock front propagates with constant velocity up to the deceleration radius $R_d\approx (E_j(\theta)/\rho\Gamma_0^2(\theta))^{1/3}$, then decelerating and following the \citet{BM76} self-similar solution\footnote{The late time emission is independent of $\Gamma_0(\theta)$ as the late self-similar dynamics depends only on the explosion energy and density of the environment.}. We assume that a distribution $N(\gamma)\propto \gamma^{-p}$ of electrons are accelerated in the shock front and  that a fraction $\epsilon_e$ of the post-shock thermal energy resides in the population of accelerated electrons while a fraction $\epsilon_B$ resides  in the post-shock magnetic field (responsible for the synchrotron radiation).  
At a fixed time (measured in the lab frame), we consider the emission of the post-shock region, assigning it to the corresponding time bin in the observer frame \citep{decolle12}.  
We caution that these results are not perfectly accurate at late times when the flow becomes mildly-relativistic and lateral expansion strongly affects the flow dynamics. Yet, these calculations  clearly illustrate the effects of changing the jet structure on the early time emission.

Figure \ref{fig5} shows light curves at 3 GHz. At $\theta_{\rm obs} \approx  0^\circ$ (i.e., for on-axis jets), the emission is dominated by the highly beamed jet core and the cocoon emission is negligible at all times. Thus, all models produce very similar light curves. For jets with lower luminosity, the cocoon emission can become dominant at late times. 
As expected, differences between the models are more evident for off-axis light curves (at $\theta_{\rm obs} \gtrsim \theta_j$). At early time the observer detects emission coming directly from the cocoon, as the core of the jet is beamed away from the observer. Then, the early time ($\approx1-10$ days in the figure) flux is much larger in structured jets than in top hat jets. While top-hat jets show the flux increasing as $t^3$, the increase is slower for Gaussian and power-law jets\footnote{The $F(t)\propto t^3$ increase is flatter when considering the lateral expansion, see \citet{Gill2019}.}. At later times (i.e. at the peak of the light curves, see Figure \ref{fig5}), the jet core decelerates and enters into the observer's field of view, dominating the afterglow emission. As in the on-axis case, the emission thereafter is nearly independent on the initial  structure of the jet.

Figure \ref{fig6} shows a comparison between our model and observations of the GRB 170817A (observations are taken from \citealt{Makhathini2020}). 
For a jet injected initially with a top hat structure, the amount of energy at large angles is not sufficient  to explain the observations, while the agreement is good for Gaussian and power-law jet models. Denser winds might be able to produce more extended and energetic cocoon for top-hat models which could in principle explain the observations as well. This certainly illustrates the large degeneracy present when trying to deduce the original structure of the jet from afterglow emission \citep[see also,][]{Gill2019,Takahashi_2020,Takahashi_2021}.

Finally, we notice that differences between model and observations at $t_{\rm obs} \gtrsim 300$ days might be due to the fact that we are not considering lateral expansion. 

Our analytical calculus describes well the afterglow emission which occurs in an optically thin medium. In this medium, the re-emission effects can be depreciated \citep[e.g.,][]{GranotSari_2002}. An emission calculus integrated by ray-tracing is not necessary in absence of re-emission effects that need a history of interactions. Then, one can estimate the flux (light curves) through the contribution of each cell in the domain within history. Our flux calculus evolves independently for each angle and it reduces the problem to one dimension. However, information about lateral expansion is forgotten. In future work, we will present the calculus emission from large-scale simulations which will contain information about lateral expansion and instabilities in the front of the shock.

Light curves computed for one set of microphysical parameters can be rescaled if, as in the case of GRB 170817A, they all land in the same spectral range \citep{Granot2012,vanEertenMcfadyen2012}. Given a set of parameters $E_j, n, \epsilon_e, \epsilon_b$, a flux $F_\nu$ and an observer time $t$, a new set of parameters is given by
\begin{eqnarray}
 \frac{t'}{t}&=&\left(\frac{k}{\lambda}\right)^{1/3},   \\
 \frac{F'}{F} &=& k \lambda^{(1+p)/4}\left(\frac{\epsilon_e'}{\epsilon_e}\right)^{p-1}\left(\frac{\epsilon_B'}{\epsilon_B}\right)^{(1-p)/4} \;,
\end{eqnarray}
being $E'= k E$ and $n'=\lambda n$. Inverting this equations, we get
\begin{eqnarray}
  n &=& n_0 \left(\frac{\epsilon_e}{0.01}\right)^{4\frac{1-p}{p+5}} \left(\frac{\epsilon_B}{0.001}\right)^{\frac{p-1}{p+5}}\;, \\
  E &=& E_0 \left(\frac{\epsilon_e}{0.01}\right)^{4\frac{1-p}{p+5}} \left(\frac{\epsilon_B}{0.001}\right)^{\frac{p-1}{p+5}}\;, 
\end{eqnarray}
with $n_0= 1.7\times 10^{-3}$ cm$^{-3}$, $5.6\times 10^{-3}$ cm$^{-3}$ and $1.5\times 10^{-2}$ cm$^{-3}$ for the models shown in Figure \ref{fig6} (from top to bottom panel), and where $E_0= 2.9 \times 10^{50}$ erg, $3.6 \times 10^{50}$ erg and $4.2 \times 10^{50}$ erg.

\section{Conclusions}\label{sec:conclusions}

In this paper we presented numerical simulations of relativistic jets associated with SGRBs. We explored different initial jet structures, duration and luminosities. The numerical simulations show that the initial jet structure plays a pivotal role on its final structure once it breaks out from the dense wind medium  expected  around the NS merger at the time of jet triggering. As such, the initial jet structure, which needs to be deduced from global simulations of neutrino-cooled disks, impacts the jet afterglow emission observed in SGRBs.

We show that high luminosity structured jets can have significantly more energy at large observing angles than jets injected with a top-hat structure. The final distribution of the jet depends on the density stratification of the environment and on the jet duration, with the initial jet structure better preserved in tenuous  media and for long-lasting jets. We also show that the afterglow emission strongly depends on the initial jet structure. Although a large degeneracy is observed  in the determination of the physical parameters, our results have highlighted the importance of understanding the initial structure of jet given its expected imprint in  observations of  afterglow emission from SGRBs.


\section*{Acknowledgements}
\addcontentsline{toc}{section}{Acknowledgements}
We thank Diego L\'opez-C\'amara, Leonardo Garc\'ia-Garc\'ia and Alan Watson for useful discussions. We also thank the anonymous referee for helpful comments on this manuscript. We acknowledge the computing time granted by DGTIC UNAM on the supercomputer Miztli (project LANCAD-UNAM-DGTIC-281). GU acknowledges support from a CONACyT scholarship for doctoral studies. GU and FDC acknowledge support from the UNAM-PAPIIT grant AG100820. AMB acknowledges support from a CONACyT-UC-Mexus doctoral scholarship. ERR and AMB thank the Heising-Simons Foundation, the Danish National Research Foundation (DNRF132), and the NSF (AST-1911206, AST-1852393, and AST-1615881) for support.
\section*{Availability of data}

The data underlying this article will be shared on reasonable request to the corresponding author.

\bibliographystyle{mnras}
\bibliography{biblio} 


\appendix

\section{On the ``plug'' instability in two-dimensional jet simulations}


Two-dimensional simulations are prone to a numerical instability known as the  ``plug'' instability.  
The plug instability is an artifact that appears frequently in simulations of relativistic and non-relativistic jets, and corresponds to a large accumulation of material at the shock front. This instability is not numerical. Indeed, it represents the correct numerical solution to a 2D problem, but not the correct physical solution.
This instability is not present in three dimensional simulations when the symmetry with respect to the direction of propagation of the jet is broken \citep[e.g.,][]{lopezcamara2013,lazzati15,lopezcamara2016}.  
In 3D simulations, the extra degree of freedom makes possible for the material in front of the shock to move laterally, while this is more difficult in 2D simulations

\begin{figure}
    \centering
    \includegraphics[scale=0.25]{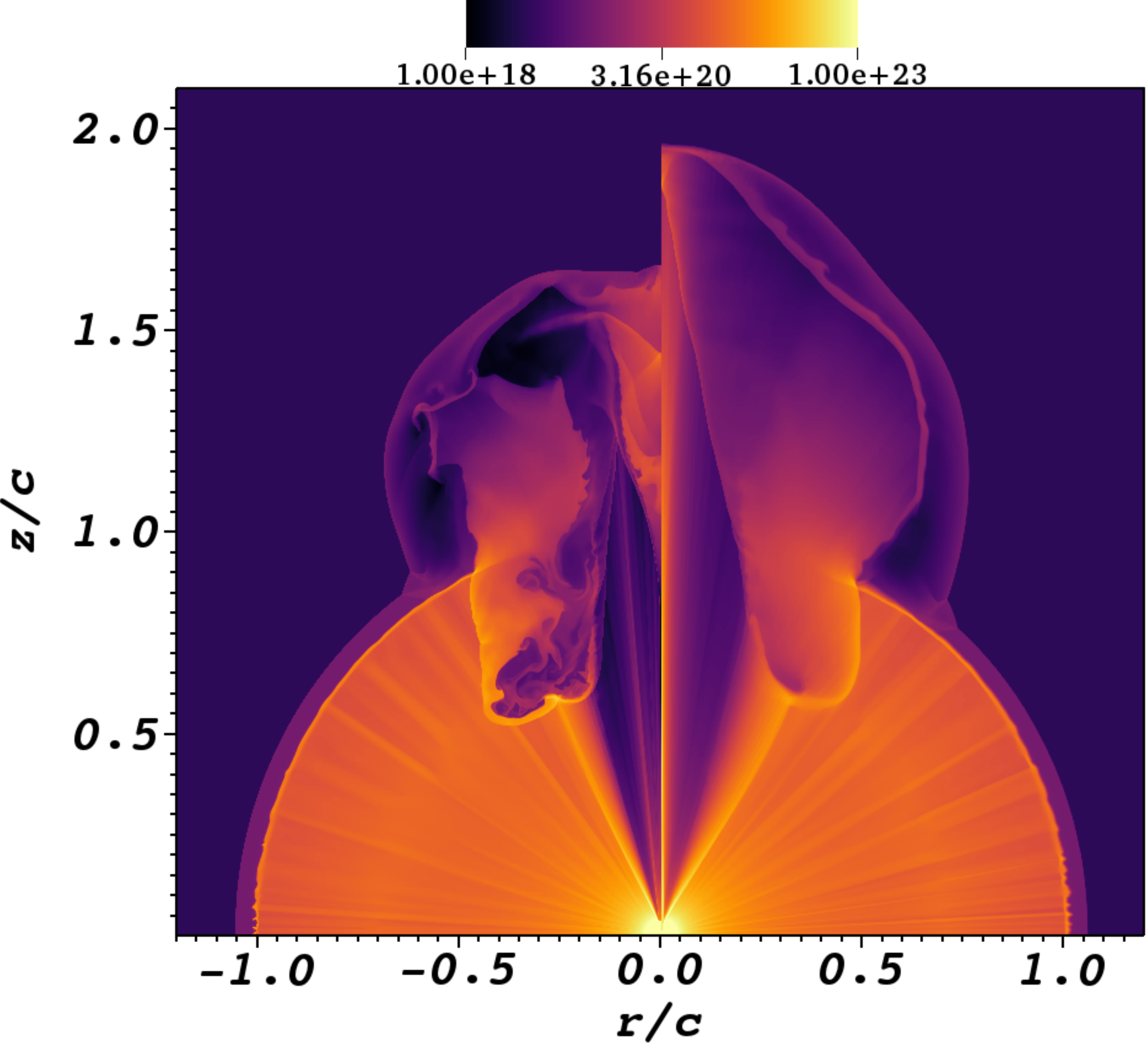}
    \caption{Density map. \emph{Left}: Jet with a top-hat, radial injection velocity. The ``plug'' instability forms an extended, dense region close to the $r=0$ axis. \emph{Right}: jet simulations done by changing the direction of the jet velocity at the inner boundary (see the text for details). The plug instability is largely suppressed.}
    \label{fig7}
\end{figure}

In our 2D simulations in cylindrical coordinates, the jet velocity in the top-hat jet is defined at the inner boundary as
\begin{equation}
    v_r = v_j \frac{r-r_0}{R} \qquad v_z = v_j \frac{z}{R} \;,
\end{equation}
being $R=\sqrt{r^2+z^2}$. When setting $r_0 = 0$, the jet velocity is radial and the simulation presents the ``plug'' instability. By implementing a small change into the direction of propagation of the jet (i.e. by assuming that the jet velocity is not perfectly radial), this instability is  suppressed (see Figure \ref{fig7}).
Substantially, this is equivalent to considering a jet with a velocity intermediate  between radial and cylindrical (with $v = v_z$), i.e. making the jet ``sharper''.

Specifically, we choose 
\begin{equation}
    r_0 = r e^{-\left(\frac{r}{\eta r_w}\right)^2}
\end{equation}
being $r_w$ the radius of the inner boundary ($=10^9$ cm in our case) and $\eta < 1$ a small value (we choose, somehow arbitrarily, $\eta=1/3$), such that the velocity is approximately radial at $r=0$ and $r=r_w$, but it slightly differs from radial otherwise.

\bsp	
\label{lastpage}
\end{document}